\documentclass[aps,superscriptaddress,prl,showpacs,twocolumn]{revtex4-2}
\usepackage{amsmath,amssymb,mathtools}

\usepackage{commath}
\usepackage{bm,bbm}
\usepackage{physics}
\usepackage{tikz}
 \usepackage{booktabs}
\usepackage[colorlinks = true,
            linkcolor = blue,
            urlcolor  = blue,
            citecolor = blue,
            anchorcolor = blue]{hyperref}
\usepackage{graphicx}
\usepackage{xcolor}
\usepackage{epstopdf}
\usepackage{CJKutf8}
\usepackage{orcidlink}

\begin{document}
    \title{
 Weighted Nested Commutators for Scalable Counterdiabatic State Preparation
}

\author{Jialiang Tang$^{\orcidlink{0009-0001-6420-6492}}$}
\affiliation{Instituto de Ciencia de Materiales de Madrid ICMM-CSIC, 28049 Madrid, Spain}
\affiliation{Departamento de Física Teórica de la Materia Condensada, Universidad Autónoma de Madrid, Madrid, Spain}

\author{Xi Chen$^{\orcidlink{0000-0003-4221-4288}}$}
\email{xi.chen@csic.es}
\affiliation{Instituto de Ciencia de Materiales de Madrid ICMM-CSIC, 28049 Madrid, Spain}

\author{Zhi-Yuan Wei \begin{CJK}{UTF8}{gbsn}(魏志远)\end{CJK}$^{\orcidlink{0000-0003-4465-2361}}$}
\email{zywei@umd.edu}
\affiliation{
Joint Quantum Institute and Joint Center for Quantum Information and Computer Science, NIST/University of Maryland, College Park, Maryland 20742, USA}
    
\begin{abstract}
Counterdiabatic (CD) driving enables efficient quantum state preparation, but it requires implementing highly nonlocal adiabatic gauge potentials (AGP) that are impractical to compute and realize in large many-body systems. We introduce a \textit{weighted nested-commutator} (WNC) ansatz to approximate AGP using local operators. The WNC ansatz generalizes the standard nested-commutator ansatz by assigning independent variational weights to commutators of local Hamiltonian terms, thereby enlarging the variational space while preserving a fixed operator range. 
We show that the WNC ansatz can be efficiently optimized using a local optimization scheme. Moreover, it systematically outperforms the nested-commutator ansatz in preparing one-dimensional matrix product states (MPS) and the ground state of a nonintegrable quantum Ising model.
We then numerically demonstrate that CD driving based on the WNC ansatz significantly accelerates the preparation of 1D MPS for system sizes up to $N = 1000$ qubits, as well as the two-dimensional Affleck–Kennedy–Lieb–Tasaki state on a hexagonal lattice with up to $N = 3 \times 10$ sites.
\end{abstract}

\maketitle

\textit{Introduction.---}
Adiabatic state preparation is a standard route to many-body ground states with broad applications in quantum simulation~\cite{RevModPhys.86.153,Daley2022PracticalQuantumAdvantage}, quantum chemistry~\cite{RevModPhys.92.015003}, and quantum optimization algorithms~\cite{RevModPhys.90.015002,PRXQuantum.3.020347,Abbas2024ChallengesQuantumOptimization}. While demonstrated across diverse scenarios~\cite{RevModPhys.90.015002,Menchon-Enrich_2016,RevModPhys.89.015006,koch2022quantum}, it typically requires long runtimes to suppress diabatic excitations, challenging near-term devices with limited coherence. To address this issue, shortcuts to adiabaticity~\cite{PhysRevLett.104.063002,RevModPhys.91.045001,j8c7-v2hd} have been developed to prepare the target state on substantially shorter timescales. 

Counterdiabatic (CD) driving~\cite{berry2009transitionless,PhysRevLett.105.123003,PhysRevLett.111.100502,PhysRevApplied.15.024038,PhysRevLett.133.123402} achieves this acceleration by adding an adiabatic gauge potential (AGP)~\cite{KOLODRUBETZ20171}, that cancels nonadiabatic transitions. However, ideal CD driving requires the exact AGP, which generally depends on full eigenspectrum information and becomes highly nonlocal in many-body systems. 
This challenge has motivated variational constructions restricted to experimentally accessible local operators~\cite{sels2017minimizing,PRXQuantum.4.010312}, as well as universal schemes based on Krylov space and performance guarantees~\cite{pqhl-nbtk,wbbs-s8fs}.
The nested-commutator (NC) expansion~\cite{PhysRevLett.123.090602,bhattacharjee2023lanczosapproachadiabaticgauge,PhysRevX.14.011032} is a widely used ansatz for the AGP, built from a hierarchy of commutators between the Hamiltonian and its parameter derivatives, 
and makes CD implementations feasible~\cite{Du2016StimulatedRamanSTAP,an2016shortcuts,PhysRevApplied.16.034050,PhysRevApplied.13.044059,Zhou_2024,Guan_2025}.

A bottleneck of the NC ansatz is that, at any fixed truncation order, it contains only a small number of variational coefficients. This limits its expressive power, making it increasingly difficult to approximate the exact AGP as the system size grows. Improving the approximation therefore requires increasing the truncation order, which rapidly enlarges the spatial support of the resulting operators and raises both the classical optimization cost and the experimental implementation overhead \cite{PhysRevA.107.022607,PhysRevB.110.024304}. 
As a result, CD benchmarks for generic non-integrable many-body systems are often limited to system sizes accessible by exact diagonalization ($N \lesssim 20$ qubits).
 \begin{figure}[t]
\centering
\includegraphics[width=\linewidth]{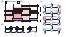}
\caption{(a) 1D geometrically local Hamiltonian (blue, illustrated for the nearest-neighbor case) [Eq.~\eqref{eq:k_general}] and the corresponding first-order ($\ell = 1$) WNC ansatz (yellow) [Eq.~\eqref{eq:wnc}]. In the global optimization scheme, all coefficients $\{\alpha\}$  are optimized simultaneously, whereas in the local optimization scheme the coefficients within local regions $\{{\cal B}_\mu\}$ (red boxes) are optimized independently. (b,c) PEPS [Eq.~\eqref{psi_f_targ}] are constructed by contracting local tensors $\{Q_v\}$ (blue circles) with maximally entangled virtual-qudit pairs (connected dots), for both a 1D system of size $N=2N_p$ [panel (b)] and a 2D hexagonal lattice [panel (c)], illustrated here with size   $N=3\times 5$. The red boxes indicate the corresponding parent Hamiltonian terms [Eq.~\eqref{eq:parent_hamiltonian}].}
\label{fig:model}
\end{figure}

\begin{figure}[t]
\centering
\includegraphics[width=7.5cm]{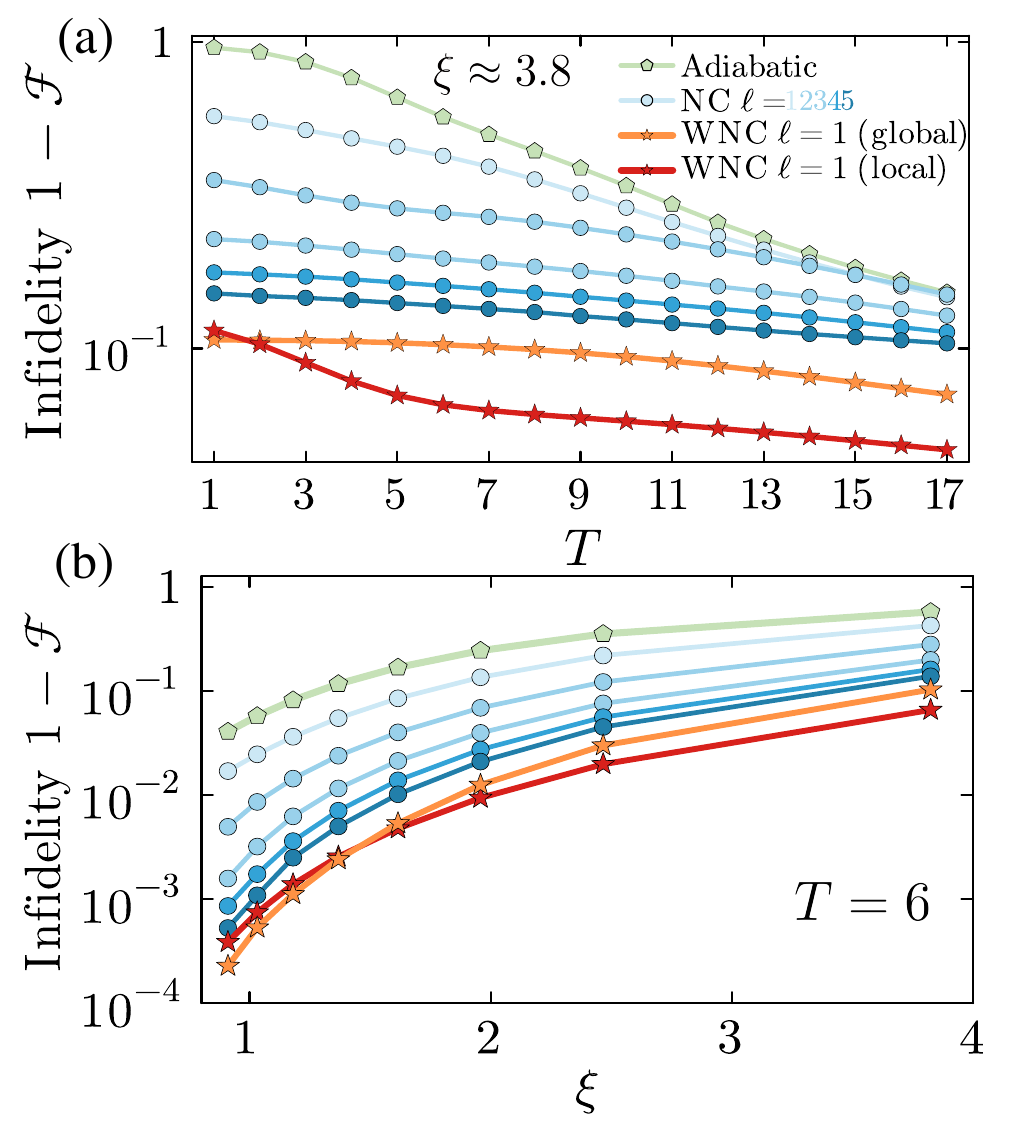}
\caption{Preparation of the MPS family [Eq.~\eqref{cls_mps}] of $N=30$ using adiabatic evolution and CD driving based on the NC and WNC ansatze. (a) Infidelity as a function of the total evolution time $T$ for a fixed target-state correlation length $\xi \approx 3.8$. (b) Infidelity as a function of $\xi$ for a fixed $T=6$. }
\label{fig:MPS_withAGP}
\end{figure}

\begin{figure}[t]
\centering
\includegraphics[width=\linewidth]{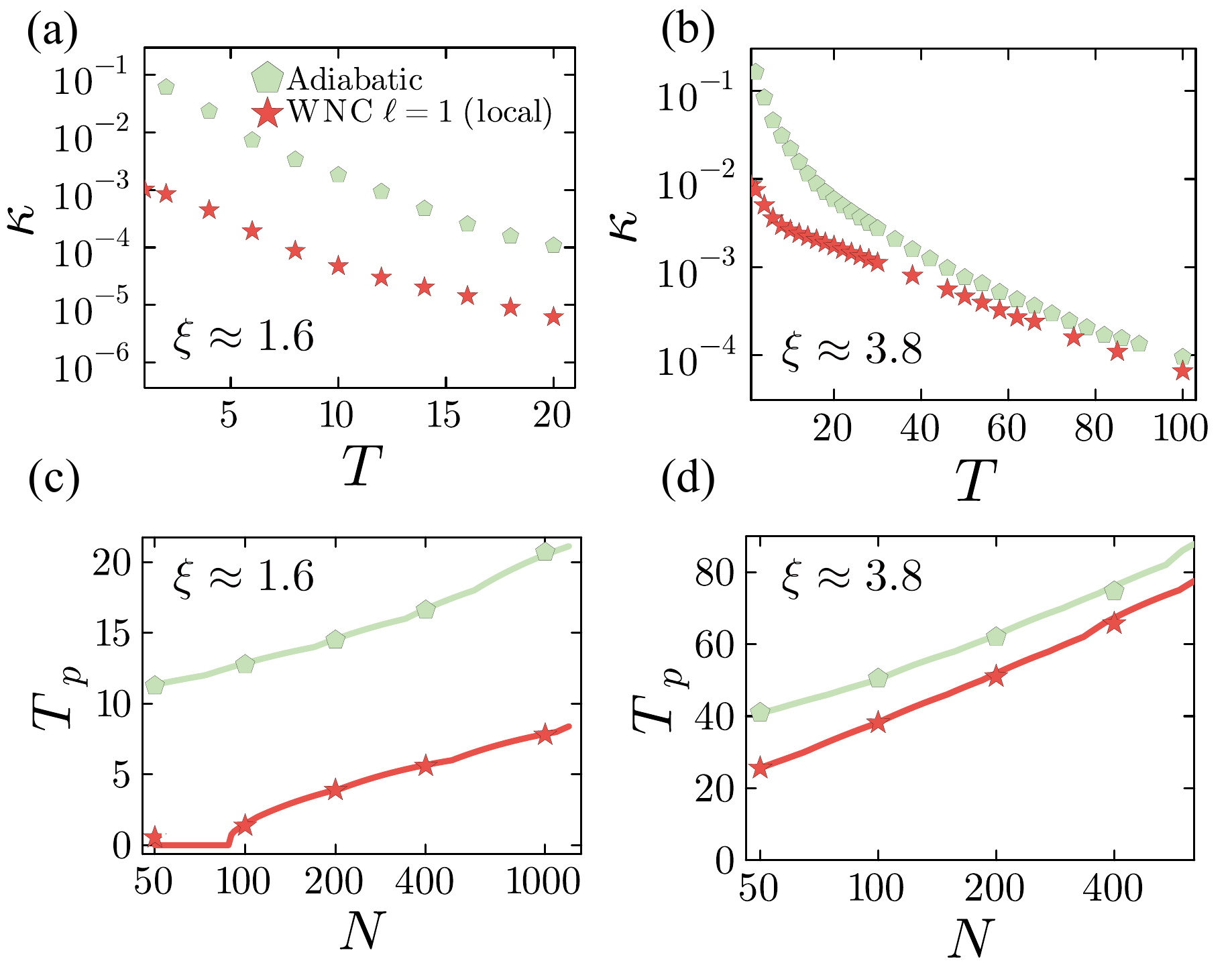}
\caption{(a),(b) Extracted error density $\kappa(T)$ [cf.~Eq.~\eqref{Eq:fid_n_scale}]  versus the evolution time $T$ for the preparation of the MPS family with correlation lengths $\xi\approx 1.6,3.8$.
(c),(d) Runtime $T_p$ required to reach the target fidelity $\mathcal{F}=0.95$ for the 1D MPS family  preparation as a function of system size $N$.
Solid green (adiabatic) and red (WNC) lines show the prediction from Eq.~\eqref{Eq:fid_n_scale}, and symbols denote TDVP simulation results. 
}
\label{fig:kappa}
\end{figure}

\begin{figure}[t]
\centering
\includegraphics[width=\linewidth]{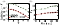}
\caption{(a)
Infidelity of preparing the 2D AKLT state on the hexagonal lattice ($N=3\times3$) as a function of the total evolution time $T$.
(b) The evolution time \( T_p \) required to prepare the 2D AKLT state on a hexagonal lattice with fidelity \( \mathcal{F} = 0.99 \) for system sizes \( N = 3 \times L \) $(L=3-10)$ for adiabatic evolution and the CD driving based on the WNC ansatz ($\ell =1$). 
}
\label{fig:2d_aklt}
\end{figure}

To overcome this limitation, we introduce a weighted nested-commutator (WNC) ansatz that preserves the NC hierarchy while assigning independent variational coefficients to local commutator-string terms [Fig.~\ref{fig:model}(a)]. 
This yields an extensive number of variational coefficients that scale with the system size $N$, thereby enlarging the variational space compared to the NC ansatz at any fixed truncation order and operator range. We show that the WNC ansatz can be efficiently optimized using a local optimization scheme. Already at first order, WNC achieves lower infidelity than high-order NC protocols for preparing a family of matrix product states (MPS) of bond dimension $D=2$ [Fig.~\ref{fig:MPS_withAGP}], as well as the ground state of the nonintegrable quantum Ising model with both transverse and longitudinal field (see the \textit{End Matter}).
We further demonstrate numerically that WNC provides scalable and substantial acceleration over adiabatic state preparation for the family of 1D MPS up to system size $N = 1000$ [Fig.~\ref{fig:kappa}], as well as for the 2D AKLT model on a hexagonal lattice up to $N = 3 \times 10$ [Fig.~\ref{fig:2d_aklt}]. These results highlight the effectiveness of WNC-based counterdiabatic driving for large many-body systems.

\textit{Model and CD driving.---}
\label{sec: Counterdibatic driving} 
We consider the quasi-adiabatic preparation of the ground state $|\psi(1)\rangle$ of a geometrically local Hamiltonian $H(1)$ on a finite-dimensional lattice of $N$ qubits~\cite{RevModPhys.90.015002} with evolution time $T$.
The system is initialized in the easy-to-prepare ground state $|\psi(0)\rangle$ of a trivial local Hamiltonian $H(0)$. 
With schedule function $s=s(t)$, we construct a geometrically local path $H(s)$ with $s(0)=0$ and $s(T)=1$. We can express the path as a sum of $N_h \sim N$ geometrically local terms $\{h_j(s)\}$, as
\begin{equation}
H(s)=\sum_{j=1}^{N_h\sim N} h_j(s).
\label{eq:k_general}
\end{equation}

The CD driving accelerates adiabatic evolution by adding the AGP ${\cal A}(s)$ that suppresses unwanted diabatic transitions. The resulting Hamiltonian reads~\cite{KOLODRUBETZ20171}
\begin{equation}
H_{\rm CD}(s) = H(s) + \dot{s}(t) \mathcal{A}(s),
\label{eq:hamiltonian_with_cd}
\end{equation}
For notational simplicity, we henceforth omit the explicit $s(t)$ dependence of $H$, $H_{\rm CD}$, and $\mathcal{A}$ when unambiguous.  The exact AGP satisfies~\cite{KOLODRUBETZ20171}
\begin{equation}  
\partial_s {H} + \frac{i}{\hbar} [\mathcal{A}, {H}] = 0.
\label{eq:agp_commucator} 
\end{equation}  

In generic interacting many-body systems, solving Eq.~\eqref{eq:agp_commucator} requires full eigenspectrum information and typically yields highly nonlocal operators, making the exact AGP impractical to compute and experimentally realize.

A widely used approach is to approximate the AGP using a truncated series of nested commutators~\cite{PhysRevLett.123.090602}. At the truncation order $\ell$, we have
\begin{equation}
    \mathcal{A}_{\rm NC}^{(\ell)} = i \sum_{k=1}^\ell \alpha_k^{\rm NC}
\underbrace{[H,[H,\ldots,[H}_{2k-1},\partial_s H]]],
\label{eq:nc}
\end{equation}
where the parameters $\{\alpha_k^{\rm NC}\}$ can be obtained variationally~\cite{sels2017minimizing,PhysRevLett.123.090602} or via frequency-domain fitting~\cite{pqhl-nbtk,wbbs-s8fs}. 
The NC series is systematic and recovers the exact AGP as $\ell\to\infty$~\cite{PhysRevLett.123.090602}. 
Increasing $\ell$ improves the accuracy but rapidly expands the operator support of each local term in $\mathcal{A}_{\rm NC}^{(\ell)}$, thereby increasing the associated optimization and implementation overhead of the method. Therefore, in practice, both the truncation order $\ell$ and the number of variational coefficients are typically small constants. This limited number of variational coefficients restricts the capability of the NC ansatz to approximate the exact AGP for large many-body systems.

\textit{WNC ansatz.---}
To enhance the variational power of the NC ansatz, we introduce the \textit{weighted nested-commutator} (WNC) ansatz by assigning independent coefficients to local nested commutator terms within the NC ansatz, getting
\begin{align}
\mathcal{A}_{\rm WNC}^{(\ell)}
=i\sum_{k=1}^{\ell}\sum_{j_0...j_{2k-1}=1}^{N_h\sim N}\alpha_{j_0...j_{2k-1}}\,
\mathcal{A}_{j_0...j_{2k-1}},
\label{eq:wnc}
\end{align}
with variational coefficients $\left\{ \alpha_{j_0 \dots j_{2k-1}} \right\}$, and the nested commutators of local terms
\begin{equation} \label{local_nc_string}
\mathcal{A}_{j_0...j_{2k-1}}=\,
\underbrace{[h_{j_{2k-1}},[h_{j_{2k-2}},\ldots,[h_{j_1}}_{2k-1},\partial_s h_{j_0}]]].
\end{equation}
We illustrate the 1D WNC ansatz for $\ell=1$ in Fig.~\ref{fig:model}(a).

Using the local decomposition of the Hamiltonian [Eq.~\eqref{eq:k_general}], it is easy to see that the NC ansatz [Eq.~\eqref{eq:nc}] is a special case of the WNC ansatz by choosing $\{ \alpha_{j_0 \dots j_{2k-1}} \}_{\forall j_0...j_k} = \alpha_k^{\rm NC}$ for all $k=1,...,\ell$. Moreover, due to the geometric locality of the system, we expect that only $O(N)$ terms within $\{\mathcal{A}_{j_0\dots j_{2k-1}}\}$ are non-vanishing. Nevertheless, we retain all terms formally in  Eq.~\eqref{eq:wnc} for generality. At each fixed order $\ell$, the WNC ansatz contains the same set of local nested commutator terms as the NC ansatz; therefore, the two have comparable costs for quantum implementation. However, the WNC ansatz involves significantly more variational coefficients than the NC ansatz, thereby systematically enhancing its variational expressiveness.

\textit{Variational optimization.---}
An established approach~\cite{sels2017minimizing} to determine the variational coefficients (denoted as a vector $\vec{\alpha}$) within the WNC ansatz $\mathcal{A}_{\rm WNC}^{(\ell)}(\vec{\alpha})$ is to minimize the Hilbert-Schmidt norm
\begin{equation}
S(\vec{\alpha})=\left\|\partial_s H+\frac{i}{\hbar}\,[\mathcal{A}_{\rm WNC}^{(\ell)}(\vec{\alpha}),H]\right\|^2,
\label{eq:action}
\end{equation}
which is equivalent in spirit to solving Eq.~\eqref{eq:agp_commucator}. 
Introducing a collective index $\eta$ to label all elements within $\vec \alpha$, we formally rewrite
$\mathcal{A}^{(\ell)}_{\rm WNC}(\vec{\alpha}) = \sum_{\eta} \alpha_{\eta} \mathcal{A}_{\eta}$.
The optimal coefficients $\vec\alpha_{\rm opt}$ are obtained from the condition $\partial_{\vec{\alpha}} S(\vec{\alpha})|_{\vec \alpha = \vec\alpha_{\rm opt}} = 0$, which yields a linear system
\begin{equation} \label{lin_eq}
G \vec{\alpha}_{\rm opt} = \vec{b},
\end{equation}
with matrix elements~\cite{SM}
\begin{equation}
G_{\eta\eta'}=\mathrm{Tr}(C_{\eta}^\dagger C_{\eta'}),\qquad b_{\eta}=-\mathrm{Tr}(C_{\eta}^\dagger \partial_s H),
\label{eq:G_trace}
\end{equation}
where $C_{\eta}=i[\mathcal{A}_{\eta},H]$. For geometrically local Hamiltonians on finite-dimensional lattices, commutators between terms that do not spatially overlap vanish, thus each matrix element in Eq.~\eqref{eq:G_trace} can be evaluated in a time independent of the system size $N$. Moreover, the number of non-vanishing terms [cf.~Eq.~\eqref{local_nc_string}] in the WNC ansatz, i.e., the dimension of the linear system Eq.~\eqref{lin_eq}, scales as $O(N)$. Therefore, the optimal coefficients $\vec{\alpha}_{\rm opt}$ can be obtained with a computational cost $O(N^3)$~\cite{SM}. We refer to this procedure as the \textit{global optimization}.

To further reduce the optimization cost, we propose a \textit{local optimization} strategy. Specifically, we partition the system into $O(N)$ geometrically local regions $\{{\cal B}_\mu\}$ that cover the full lattice [cf.~Fig.~\ref{fig:model}(a)]. For each region ${\cal B}_\mu$, we define $H_{{\cal B}_\mu}$ as the sum of all Hamiltonian terms in $H$ [Eq.~\eqref{eq:k_general}] that are supported within ${\cal B}_\mu$. Similarly, we construct $\mathcal{A}_{\rm WNC}^{(\ell,{\cal B}_\mu)}$ by summing the nested commutators of local terms supported in ${\cal B}_\mu$. We then construct the action in the same manner as in Eq.~\eqref{eq:action}, but with the Hamiltonian and the WNC ansatz replaced by their local counterparts, $H_{{\cal B}_\mu}$ and $\mathcal{A}_{\rm WNC}^{(\ell,{\cal B}_\mu)}$. Applying the same procedure as in Eqs.\eqref{lin_eq} and \eqref{eq:G_trace}, we obtain the coefficients associated with the WNC terms supported within ${\cal B}_\mu$. Scanning over $\{{\cal B}_\mu\}$ that cover the full lattice, we determine all coefficients. Note that, certain coefficients will get different locally optimized values due to the potential overlap of the local regions $\{{\cal B}_\mu\}$ [e.g. the term $\alpha_{22} {\cal A}_{22}$ in Fig.~\ref{fig:model}(a)], and we average these values. The local optimization approach reduces the scaling of the computational overhead from $O(N^3)$ to $O(N)$~\cite{SM}.

\textit{PEPS and adiabatic path.---} 
In the remainder of this work, we apply CD driving based on the WNC ansatz to accelerate the adiabatic preparation of projected entangled pair states (PEPS), following the same adiabatic path studied in Ref.~\cite{PhysRevResearch.5.L022037}. In the following, we briefly review PEPS and the associated adiabatic path~\cite{PhysRevResearch.5.L022037}.

We consider a PEPS on a graph $\mathcal{G}$, with edges $\mathcal{E}$ and vertices $\mathcal{V}$, written as   
\begin{equation} \label{psi_f_targ}
\left|\psi_{\rm PEPS}\right\rangle \propto \bigotimes_{v \in \cal V} Q_{v} \bigotimes_{e \in \cal E} \left| \Phi^+ \right\rangle_e,
\end{equation}
where \( Q_v \) acts on the maximally entangled state \(|\Phi^{+}\rangle _e\propto \sum_{\alpha=0}^{D-1} |\alpha\,\alpha\rangle
\) with bond dimension $D$ of PEPS [cf.~Fig.~\ref{fig:model}(b,c)].  The MPS can be viewed as 1D PEPS with 2 virtual qudits per site, with \( Q_v \) in the bulk takes the following form:
\begin{equation} \label{proj_qk}
Q_v^{\rm 1D} = \sum\limits_{i_v = 0}^{d-1} \sum\limits_{\alpha_v ,\beta_v = 0}^{D-1} A_{[v] \alpha_v \beta_v }^{i_v} \left| i_v \right\rangle \left\langle \alpha_v \beta_v \right|,
\end{equation}
where \( A_{[v]}^{i_v} \) represents the MPS tensors. If the $\{Q_v\}$ are left-invertible, a PEPS is called \textit{injective}.

The first example we study is the \(D=2\) MPS family [cf.~Fig.~\ref{fig:model}(b)] introduced in Ref.~\cite{PhysRevLett.97.110403}, and subsequently used as a benchmark in various studies~\cite{PhysRevResearch.5.L022037,PhysRevResearch.4.L022020,PRXQuantum.5.030344,PhysRevLett.132.040404,wei2025statepreparationparallelsequentialcircuits,scheer2025renormalizationgroupbasedpreparationmatrixproduct,fischer2025largescaleimplementationquantumsubspace}. 
By grouping neighboring pairs within the bulk of $N=2N_p$ qubits into $N_p$ qudits of dimension $d=4$ [cf.~Fig.\ref{fig:model}(b)], we obtain an injective MPS with physical dimension \(d = D^2 = 4\). The corresponding tensor in Eq.~\eqref{proj_qk} is
\begin{align} \label{cls_mps}
A_{[v]}^{0}(g) &= \begin{pmatrix}
0 & 0\\
1 & 1
\end{pmatrix}, \quad 
A_{[v]}^{1}(g) = \begin{pmatrix}
0 & 0\\
1 & g
\end{pmatrix}, \\
A_{[v]}^{2}(g) &= \begin{pmatrix}
g & g\\
0 & 0
\end{pmatrix}, \quad 
A_{[v]}^{3}(g) = \begin{pmatrix}
1 & g\\
0 & 0
\end{pmatrix}. \nonumber
\end{align}
We focus on $g<0$, where this MPS family exhibit symmetry-protected topological (SPT) order, interpolating between the cluster state at $g=-1$ and the GHZ state at $g=0$~\cite{PhysRevResearch.4.L022020,PhysRevLett.97.110403}. By tuning $g$, one can control the correlation length of the state as $\xi=\left|\ln\!\left(\frac{1-g}{1+g}\right)\right|^{-1}$.

Another example we consider is the 2D AKLT state on a hexagonal lattice with cylinder boundary condition [cf.~Fig.~\ref{fig:model}(c)], which exhibits weak SPT order~\cite{you2014wave} and serves as a universal resource for measurement-based quantum computation~\cite{Wei2011}. The AKLT state can be expressed as an injective PEPS with bond dimension $D = 2$~\cite{SM}. It can be prepared deterministically using adiabatic algorithms~\cite{PhysRevLett.108.110502,PhysRevLett.116.080503,PhysRevResearch.5.L022037}, or probabilistically via measurement-based protocols~\cite{Murta2022,guo2026measurement}.

To get an adiabatic path, we construct a PEPS interpolation $|\psi(s)\rangle$ [cf.~Eq.~\ref{psi_f_targ}] with tensors $Q_v(s)=s\,Q_v+(1-s)\mathbbm{1}$ with $s\in[0,1]$, which connects the product state of maximally entangled pairs $|\psi(0)\rangle=\bigotimes_{e\in\mathcal{E}}|\Phi^+\rangle_e$ to the target state $|\psi(1)\rangle\equiv |\psi _{\rm PEPS}\rangle$.  The parent Hamiltonian of $|\psi(s)\rangle$ is~\cite{PhysRevResearch.5.L022037,PhysRevLett.97.110403,perezgarcia2007pepsuniquegroundstates} [cf.~Fig.~\ref{fig:model}(b)]
\begin{equation}
H(s)=\sum_e \Pi_{\rm ker}\!\left[\rho_e(s)\right],
\label{eq:parent_hamiltonian}
\end{equation}
where each term $\Pi_{\rm ker}[\rho_e(s)]$ denotes the projector onto the kernel of the reduced density matrix $\rho_e(s)$ on edge $e$, and will be identified as a local term in Eq.~\eqref{eq:k_general}.

For all examples, the time-dependent variational principle (TDVP) is used for numerical simulation~\cite{SM}. We compute the fidelity $\mathcal{F}=|\langle\psi(1)|\phi(T)\rangle|^2$ between the evolved state $|\phi(T)\rangle$ for time $T$ and the target ground state $|\psi(1)\rangle$. We adopt the schedule function $s_{\rm 1D}(t)=\sin^2\!\left[\frac{\pi}{2}\sin^2\!\left(\frac{\pi t}{2T}\right)\right]$ for the 1D MPS family, and $s(t)_{\mathrm{2D}} = \sin^2(\pi t / 2T)$ for the 2D AKLT state~\cite{PhysRevResearch.5.L022037}.

\textit{Benchmarking WNC.---}
To benchmark the WNC ansatz, we study the preparation of the 1D MPS family with system size $N=30$ using the adiabatic evolution [cf.~Eq.~\eqref{eq:k_general}] and the CD-assisted evolution [cf.~\eqref{eq:hamiltonian_with_cd}] using either the high-order NC or first-order WNC ansatz. 

For a fixed correlation length $\xi \approx 3.8$ of the target state, Fig.~\ref{fig:MPS_withAGP}(a) shows that, compared to adiabatic evolution, CD driving based on the NC ansatz systematically reduces the infidelity as the truncation order $\ell$ increases. Moreover, the WNC ansatz, based on either global or local optimization, yields a significantly lower infidelity than the NC ansatz: the first-order WNC ansatz already outperforms the fifth-order NC ansatz across the entire range of $T$.

Fixing time $T=6$, Fig.~\ref{fig:MPS_withAGP}(b) further demonstrates that the advantage of the first-order WNC ansatz over up to fifth-order NC ansatz persists across target states with different correlation lengths. In particular, within the same evolution time, the first-order WNC ansatz achieves nearly a three-order-of-magnitude reduction in infidelity compared to adiabatic evolution. The speedup is more pronounced for states with shorter correlation lengths, as the finite speed of information propagation limits how much states with longer correlations can be accelerated within a fixed evolution time. Finally, in the \textit{End Matter}, we show that the first-order WNC ansatz likewise outperforms the high-order NC ansatz in preparing the ground state of a 1D non-integrable quantum Ising model.

These results demonstrate that the additional variational degrees of freedom in the WNC ansatz significantly enhance the state-preparation performance compared to the NC ansatz. We further show in the SM that, for digital Hamiltonian simulation using first-order Trotterization, CD driving based on the first-order WNC ansatz achieves high-fidelity state preparation with far fewer gates than either adiabatic evolution or CD driving based on the high-order NC ansatz~\cite{SM}.
Moreover, the local optimization scheme is found to be both computationally more efficient and capable of achieving lower infidelity overall. In the following, we therefore adopt the WNC ansatz with local optimization to study the CD-assisted preparation of large-scale many-body ground states.

\textit{Large-scale preparation of 1D MPS.---}
For large-scale quasi-adiabatic state preparation, it has been shown that the fidelity $\mathcal{F}(N,T)$ exhibit an exponential decay with the system size $N$~\cite{PhysRevResearch.5.L022037}
\begin{equation}
\label{Eq:fid_n_scale}
\mathcal{F}(N,T) = \exp\left[-\kappa(T)\cdot N - c(T)\right],
\end{equation}
with error density $\kappa(T)$ reduces when increasing the evolution time $T$, together with a system-size independent term $c(T)$ induced by the boundary effects.

 Under CD driving based on the  WNC ansatz ($\ell=1$), the fidelity exhibits the same scaling behavior as in Eq.~\eqref{Eq:fid_n_scale}, but with a reduced error density $\kappa(T)$~\cite{SM}.
 The extracted $\kappa(T)$ for the preparation of the MPS family is presented in Fig.~\ref{fig:kappa}(a,b). For a short correlation length $\xi \approx 1.6$, CD driving based on the WNC ansatz reduces $\kappa(T)$ by nearly two orders of magnitude. For a longer correlation length $\xi \approx 3.8$, the reduction is more pronounced at short evolution times and gradually vanishes at large $T$, where the system approaches the adiabatic limit. This behavior is expected, as the derivative $\dot s(t)$ in Eq.~\eqref{eq:hamiltonian_with_cd} becomes negligible in the large-$T$ limit.

Using Eq.~\eqref{Eq:fid_n_scale}, one can predict the required evolution time $T_p$ to prepare an $N$-particle state with a desired target fidelity. In Fig.~\ref{fig:kappa}(c,d), we present the predicted $T_p$ for preparing the MPS family with $\mathcal{F}=0.95$, and compare it with TDVP simulations of the evolution for system sizes up to $N=1000$, finding excellent agreement. For example, for $\xi \approx 1.6$ and a fixed evolution-time budget of $T_p \approx 10$, CD driving based on the WNC ansatz ($\ell=1$) can prepare states of a size nearly two orders of magnitude larger than those achievable via adiabatic evolution. Additional results for $\xi \approx 0.9,2.5$, together with the extracted $c(T)$, are provided in the SM~\cite{SM}.

\textit{2D AKLT states.---} 
Going beyond 1D systems, we present numerical results for preparing the 2D AKLT state on a hexagonal lattice [cf.~Fig.~\ref{fig:model}(c)] with system size $N=3\times L$ in Fig.~\ref{fig:2d_aklt}. For a fixed size $N=3\times 3$ [panel (a)], the CD driving based on the WNC ansatz ($\ell=1$) reduces the state-preparation infidelity by more than an order of magnitude compared to adiabatic evolution across the entire range of evolution times $T$. For a target fidelity $\mathcal{F}=0.99$ [panel (b)], the CD driving based on the WNC ansatz approximately halves the required evolution time $T_p$ for lattices up to $N=3\times 10$ sites (each bulk site contains three qubits), which is well beyond the reach of exact diagonalization. 

\textit{Discussions.---}
We introduced the WNC ansatz together with a local optimization scheme for tuning its parameters, and showed that it systematically outperforms the standard NC ansatz. Our numerical results demonstrate that CD driving based on the WNC ansatz can significantly accelerate the adiabatic preparation of large interacting many-body ground states in both 1D systems (up to $N = 1000$) and 2D lattices (up to $N = 3 \times 10$, with three qubits per site).

Promising future directions include numerically exploring higher-order WNC ansatz to further enhance performance and clarify their scaling properties; designing experimental implementations of WNC on near-term quantum platforms~\cite{PhysRevLett.123.090602,PhysRevA.100.012341,PhysRevLett.133.123402}; developing analytical or semi-analytical methods for determining the WNC coefficients, potentially with provable performance guarantees~\cite{wbbs-s8fs,pqhl-nbtk}; extending the WNC ansatz to the preparation of gapless ground states~\cite{PhysRevB.72.161201,tgzt-dy3h} and long-range entangled phases~\cite{gjonbalaj2025shortcuts}; and devising hybrid acceleration strategies that combine WNC ansatz with complementary approaches such as optimal control, measurement-induced cooling~\cite{PRXQuantum.5.030301} and machine-learning-based techniques~\cite{j8c7-v2hd}.

\section*{Acknowledgments}
We thank Andrew Childs and Daniel Malz for insightful discussions. XC is supported by project Grant No. PID2021-125823NA-I00 funded by MCIN/AEI/10.13039/501100011033 and by ``ERDF A way of making Europe" and ``ERDF Invest in your Future'',  Basque Government through Grant No. IT1470-22, and the Severo Ochoa Centres of Excellence program through Grant CEX2024-001445-S. ZYW is supported in part by the NSF STAQ program.

\bibliography{bibfile}

\section{End matter: Benchmark for Quantum Ising Model}
% \textit{Benchmark for Quantum Ising Model.---}
As an additional benchmark, we apply the CD driving based on the WNC ansatz ($\ell=1$) to the non-integrable quantum Ising model with both transverse and longitudinal field,
\begin{align}
H(1) = h_z \sum_{j=1}^{N} Z_j + h_x \sum_{j=1}^{N} X_j + J \sum_{j=1}^{N-1} Z_j Z_{j+1},
\label{Eq:TFIM}
\end{align}
where $\{X_j\}$ and $\{Z_j\}$ are Pauli operators. 

We consider the following linear Hamiltonian interpolation that connects the trivial initial Hamiltonian $H(0)=\,h_z \sum_{j=1}^{N} Z_j$ to $H(1)$, as
\begin{equation}
H(s)=(1-s)\,h_z \sum_{j=1}^{N} Z_j
+s\left(h_x \sum_{j=1}^{N} X_j + J \sum_{j=1}^{N-1} Z_j Z_{j+1}\right),
\label{path:tfim}
\end{equation}
with schedule function $s(t)=\sin^2\!\left(\frac{\pi}{2}\sin^2\!\left(\frac{\pi t}{2T}\right)\right)$, and the parameters $J=1, h_x=2, h_z=1$.

At first order ($\ell=1$), the local nested commutators terms within the WNC ansatz [Eq.~\eqref{eq:wnc}] yield a minimal set of  local operators consisting of single-site $\{Y_j\}$ and two-body operators $\{Y_j Z_{j+1}\}$ and $\{Z_j Y_{j+1}\}$. Assigning independent coefficients to each local string, the first-order WNC ansatz is
\begin{equation}
\mathcal{A}_{\mathrm{WNC,Ising}}^{(\ell=1)}=
\sum_{j=1}^{N}\alpha_j^{y} Y_j
+\sum_{j=1}^{N-1}\Bigl(\alpha_j^{yz} Y_j Z_{j+1}+\alpha_j^{zy} Z_j Y_{j+1}\Bigr),
\label{Eq:A_OC}
\end{equation}
with coefficients $\{\alpha_j^{y},\alpha_j^{yz},\alpha_j^{zy}\}$. We employ the same global and local optimization schemes described in the main text to obtain the optimal coefficients. In the local scheme, the regions consist of three neighboring sites $(j,j+1,j+2)$ for $j=1,...,N-2$.

\begin{figure}[h!]
\centering
\includegraphics[width=\linewidth]{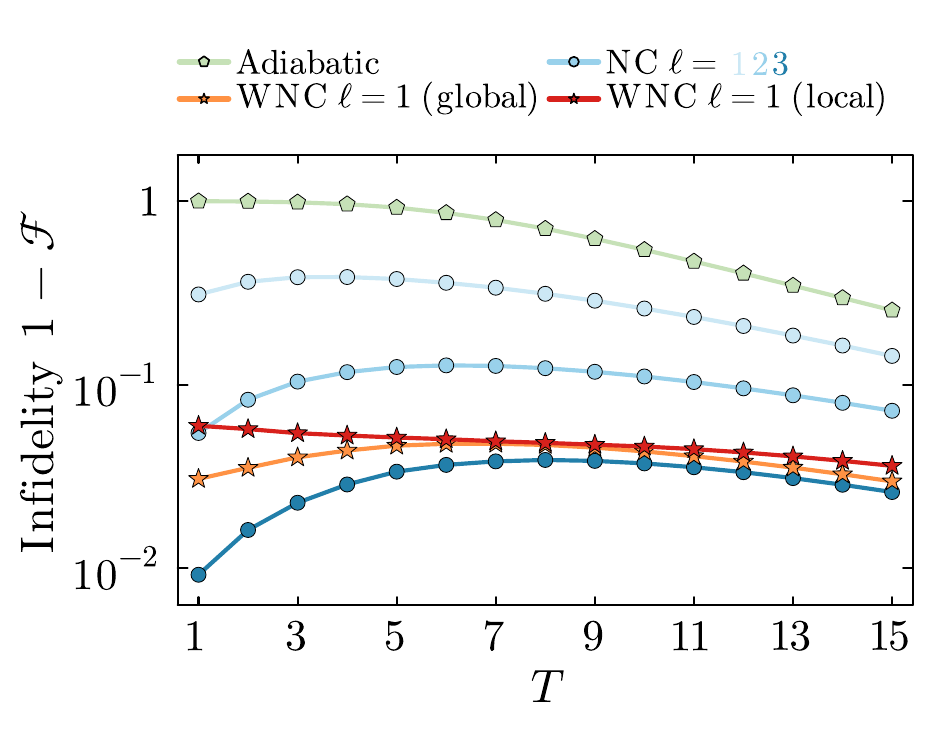}
\caption{Infidelity for preparing the ground state of~\eqref{Eq:TFIM} along the path~\eqref{path:tfim} as a function of total time $T$.
Parameters: $h_x=2$, $h_z=J=1$, $N=15$.}
\label{fig:Ising_model}
\end{figure}

We simulate the evolution along the path in Eq.~\eqref{path:tfim} for $N=15$ with $h_x=2$ and $h_z=J=1$, and compare (i) adiabatic evolution, (ii) CD driving based on the NC ansatz at truncation orders $\ell=1,2,3$, and (iii) first-order WNC with both global and local optimization.

The results are shown in Fig.~\ref{fig:Ising_model}. Similar to the preparation of the 1D MPS family [cf.~Fig.~\ref{fig:MPS_withAGP}], the first-order WNC ansatz already achieves an infidelity comparable to that of the high-order NC ansatz ($\ell=3$ in this case). Moreover, the performance of the local optimization closely matches that of the global optimization, indicating that the local scheme offers both improved optimization efficiency and competitive state-preparation performance.

\end{document}

\crefname{equation}{Eq.}{Eqs.}
\crefname{figure}{Fig.}{Fig.}
\crefname{appendix}{Appendix}{Appendix}
\renewcommand{\thefigure}{S\arabic{figure}}
\renewcommand{\theequation}{S\arabic{equation}}
\renewcommand{\bibnumfmt}[1]{[S#1]}
\renewcommand{\citenumfont}[1]{S#1}
\renewcommand{\thesection}{S\arabic{section}}

\title{
Supplementary Materials for: \\ Weighted Nested Commutators for Scalable Counterdiabatic State Preparation
}

\author{Jialiang Tang$^{\orcidlink{0009-0001-6420-6492}}$}
\affiliation{Instituto de Ciencia de Materiales de Madrid ICMM-CSIC, 28049 Madrid, Spain}
\affiliation{Departamento de Física Teórica de la Materia Condensada, Universidad Autónoma de Madrid, Madrid, Spain}

\author{Xi Chen$^{\orcidlink{0000-0003-4221-4288}}$}
\email{xi.chen@csic.es}
\affiliation{Instituto de Ciencia de Materiales de Madrid ICMM-CSIC, 28049 Madrid, Spain}

\author{Zhi-Yuan Wei \begin{CJK}{UTF8}{gbsn}(魏志远)\end{CJK}$^{\orcidlink{0000-0003-4465-2361}}$}
\email{zywei@umd.edu}
\affiliation{
Joint Quantum Institute and Joint Center for Quantum Information and Computer Science, NIST/University of Maryland, College Park, Maryland 20742, USA}
    
\maketitle
\tableofcontents

\section{Optimizing the variational coefficients}
\label{appendix:multi_param}
Here we provide more details on the optimization scheme used to optimize the parameters in the weighted nested commutator (WNC) ansatz [Eq.(5) in the main text], and discuss the associated computational cost. The same optimization procedure is also applied to determine the coefficients in the standard nested commutator (NC) ansatz [Eq.(4) in the main text] studied in this paper.

\subsection{The optimization scheme}

For the WNC ansatz, each non-vanishing nested commutator of local terms [Eq.(6) in the main text] has an independent parameter, which we collectively denote as
$\mathcal{A}^{(\ell)}_{\rm WNC}(\vec{\alpha}) = \sum_{\eta} \alpha_{\eta} \mathcal{A}_{\eta}$.
The optimal coefficients $\vec{\alpha}_{\rm opt}$ can be determined by minimizing the Hilbert-Schmidt norm~\cite{sels2017minimizing}
\begin{equation}
S(\vec{\alpha})=\left\|\partial_s H+\frac{i}{\hbar}\,[\mathcal{A}_{\rm WNC}^{(\ell)}(\vec{\alpha}),H]\right\|^2,
\label{global_action}
\end{equation}
where $\|O\|^2:=\mathrm{Tr}(O^\dagger O)$. We set $\hbar=1$ from now on.

Defining $C_{\eta}:=i[\mathcal{A}_{\eta},H]$, we can expand $S(\vec{\alpha})$ as
\begin{align}
S(\vec{\alpha})
=\mathrm{Tr}\!\left[\partial_s H^\dagger \partial_s H\right]
+2\sum_{\eta} \alpha_{\eta}\, \mathrm{Tr}\!\left(C_{\eta}^\dagger \partial_s H\right)
+\sum_{{\eta},{\eta'}}\alpha_{\eta}\alpha_{\eta'} \,\mathrm{Tr}\!\left(C_{\eta}^\dagger C_{\eta'} \right).
\label{eq:quadratic_cost}
\end{align}
Further defining the positive semi-definite Gram matrix $G$ and the vector $\vec b$, with elements
\begin{equation} \label{Gb_form}
G_{\eta\eta'}=\mathrm{Tr}(C_{\eta}^\dagger C_{\eta'})	,\qquad b_{\eta}=-\mathrm{Tr}(C_{\eta}^\dagger \partial_s H).
\end{equation}
With that, the stationarity condition $\partial_{\vec{\alpha}} S(\vec{\alpha})|_{\vec \alpha = \vec\alpha_{\rm opt}} = 0$ for determining the optimal coefficients can be written as
\begin{equation}
\frac{\partial S}{\partial \vec{\alpha}} |_{\vec \alpha = \vec\alpha_{\rm opt}}
= -2 \vec{b} + 2 G  \vec{\alpha}_{\rm opt} = 0.
\end{equation}
Thus, $\vec{\alpha}_{\rm opt}$ is obtained by solving the linear system $G \vec{\alpha}_{\rm opt} = \vec{b}$ [Eq.~(8) in the main text].

\subsection{Scaling of the optimization cost}
Assuming a total evolution time $T$ and implementing the time evolution via Trotterization with time step $\Delta t$ (see the TDVP simulation in \cref{tdvp_simu}), the optimization described in \cref{appendix:multi_param}—with the time-dependent Hamiltonian $H[s(t)]$—is performed at each of the $T/\Delta t$ steps. In the following, we analyze the scaling of the computational cost with system size $N$ for a single step for both the global optimization and local optimization schemes (see the main text for descriptions of these two schemes), focusing on geometrically local Hamiltonian path on finite-dimensional lattices.

\subsubsection{Global optimization}
\label{glob_scale}
For each step, the global optimization consists of two parts: 
(i) computing the elements of $G$ and $\vec{b}$, and 
(ii) solving the linear system $G \vec{\alpha}_{\rm opt} = \vec{b}$.

For a geometrically local Hamiltonian path, nested commutators of the local terms [Eq.~(6) in the main text] that do not spatially overlap vanish. Consequently, the total number of variational coefficients associated with the non-vanishing terms scales as $\mathrm{dim}(\vec{\alpha}) = O(N)$. Therefore, the matrix $G$ contains $O(N^2)$ elements, while the vector $\vec{b}$ contains $O(N)$ elements. 

Using the geometric locality of the Hamiltonian once again, one can show that the commutators $\{C_{\eta}\}$ are also geometrically local, and each can be evaluated in $O(1)$ time. Combining the locality of $\{C_{\eta}\}$ with \cref{Gb_form}, it follows that each element of $G$ and $\vec{b}$ can likewise be evaluated in $O(1)$ time. 

In summary, the computational cost of part (i) is dominated by evaluating the elements of $G$, which scales as $O(N^2)$. For part (ii), solving a linear system of dimension $O(N)$ generally requires $O(N^3)$ computational cost. Therefore, the overall asymptotic computational cost of the global optimization per step scales as $O(N^3)$.

\subsubsection{Local optimization}
In the local optimization scheme, we partition the system into $O(N)$ geometrically local regions $\{{\cal B}_\mu\}$ that cover the full lattice. For each region ${\cal B}_\mu$, we define $H_{{\cal B}_\mu}$ as the sum of all Hamiltonian terms in $H$ that are supported within ${\cal B}_\mu$. Similarly, we construct $\mathcal{A}_{\rm WNC}^{(\ell,{\cal B}_\mu)}$ by summing the nested commutators of local terms supported in ${\cal B}_\mu$. We then construct the action in the same manner as in \cref{global_action}, but with the Hamiltonian and the WNC ansatz replaced by their local counterparts, $H_{{\cal B}_\mu}$ and $\mathcal{A}_{\rm WNC}^{(\ell,{\cal B}_\mu)}$. 

Following the same procedure as in \cref{appendix:multi_param}, together with the scaling analysis for the global optimization case in \cref{glob_scale}, one finds that $O(N)$ individual optimizations are required, one for each local region. However, the computational cost of each local optimization scales as $O(1)$. Therefore, the total computational cost per evolution step for the local optimization approach scales as $O(N)$.

Compared to the global optimization scheme, whose single-step computational cost scales as $O(N^3)$, the local optimization approach significantly reduces this overhead to $O(N)$. 
Moreover, although the local optimization is expected to yield suboptimal solutions with respect to minimizing the global action in Eq.~\eqref{global_action}, this does not necessarily imply a lower state-preparation fidelity compared to the global optimization scheme. In fact, as shown in the main results, the local optimization achieves an infidelity comparable to that obtained via global optimization.

\section{2D AKLT state on the hexagonal lattice}

The 2D AKLT state of spin-$S=3/2$ on the hexagonal lattice can be expressed as a PEPS with bond dimension $D=2$, as illustrated in Fig.~1(a) in the main text. Each bulk operator $Q_v$ consists of a projector $P_{S,v}$ and a singlet matrix
$Y=\begin{pmatrix}
0 & -1 \\
1 & 0
\end{pmatrix},$ where $P_{S,v}$ projects the $n_v$ qubits at vertex $v$ onto their symmetric subspace, and each $Y$ transforms a virtual entangled pair into a singlet~\cite{RevModPhys.93.045003}. For the hexagonal lattice, the bulk operator takes the form
$Q_v = P_{3/2,v}\,
  (\mathbbm{1}\otimes Y \otimes Y)$.
The local projector $P_{3/2,v}$ is explicitly given by
\begin{align}
P_{3/2,v} &= |000\rangle\langle 000| + |111\rangle\langle 111| \notag \\
&\quad + \tfrac{1}{3}\,(|011\rangle+|101\rangle+|110\rangle)
  (\langle 011|+\langle 101|+\langle 110|) \notag \\
&\quad + \tfrac{1}{3}\,(|001\rangle+|010\rangle+|100\rangle)
  (\langle 001|+\langle 010|+\langle 100|).
\end{align}
Here, the virtual qubits in the AKLT construction are promoted to physical qubits. In the 2D hexagonal lattice case, each site hosts $n_v=3$ qubits, and the physical spin-$3/2$ degree of freedom corresponds to the symmetric subspace with basis states
\begin{align}
|S_z=+\tfrac{3}{2}\rangle &= |000\rangle, \\
|S_z=+\tfrac{1}{2}\rangle &= \tfrac{1}{\sqrt{3}}(|001\rangle+|010\rangle+|100\rangle), \\
|S_z=-\tfrac{3}{2}\rangle &= |111\rangle, \\
|S_z=-\tfrac{1}{2}\rangle &= \tfrac{1}{\sqrt{3}}(|011\rangle+|101\rangle+|110\rangle).
\end{align}

\begin{figure*}[b]
\centering
\includegraphics[width=15cm]{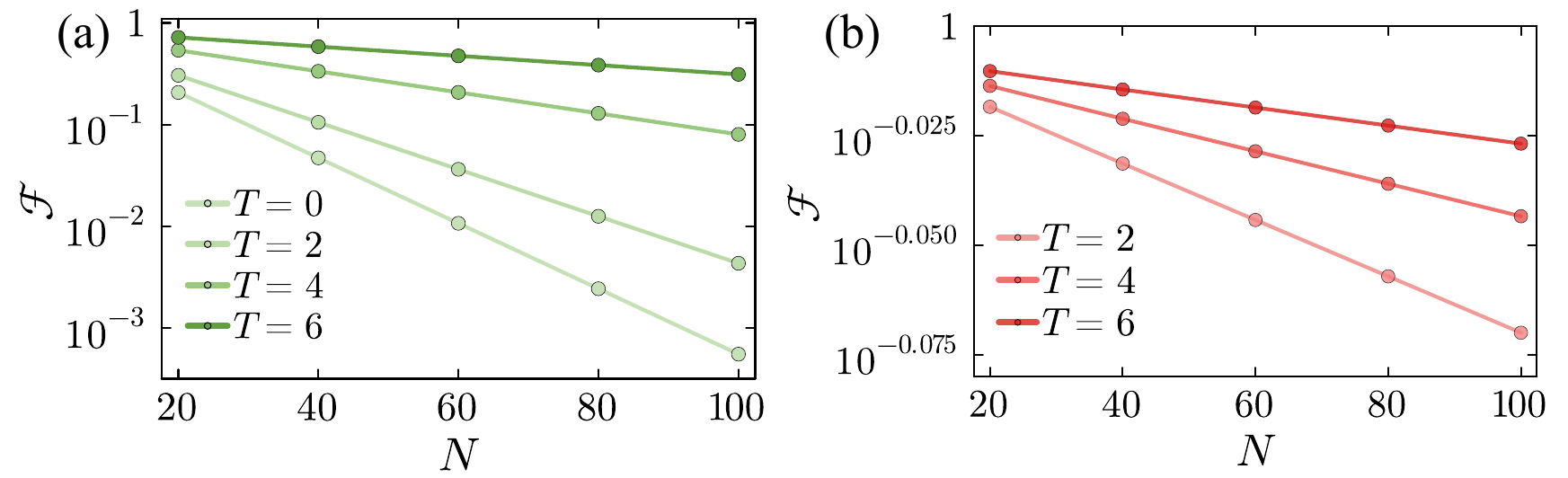}
\caption{
The fidelity $\mathcal{F}$ for preparing a 1D MPS with correlation length $\xi \approx 2.5$, shown as a function of system size $N$ for various evolution times $T$, using (a) adiabatic evolution and (b) CD driving with the first-order WNC ansatz.}
\label{fig:exponential_decay}
\end{figure*}

\section{Numerical simulation details}
\label{tdvp_simu}
We use matrix product states (MPS) to represent the quantum state during both the adiabatic and counterdiabatic dynamics. For the 2D hexagonal lattice with cylindrical boundary conditions [cf.~Fig.~1(c) in the main text], we adopt a 1D snake ordering to label the lattice sites (see the Supplementary Material of Ref.~\cite{PhysRevResearch.5.L022037}). 

The time-dependent variational principle (TDVP)~\cite{PhysRevLett.107.070601,PhysRevB.94.165116} is employed to evolve the MPS during the adiabatic and counterdiabatic dynamics. To compute $\partial_s h_j$ needed for constructing the local terms within the WNC ansatz [cf. Eq.(6) in the main text], we use a finite-difference approximation,
\begin{equation}
\partial_s h_j(s) \approx \frac{h_j(s + \Delta s) - h_j(s - \Delta s)}{2\Delta s},
\end{equation}
where $\Delta s = 0.01$ is sufficient to ensure numerical convergence.

All simulations are carried out using the TDVP implementation in the \texttt{ITensor.jl} library~\cite{itensor}. We set the single-step simulation error per MPS bond to $\delta = 10^{-9}$ (1D MPS), $\delta = 2\times10^{-8}$ (2D AKLT), and the time step to $dt = 0.05$ ($\delta$ and $dt$ are the parameters of the TDVP function within \texttt{ITensor.jl}), which are sufficient to obtain converged results.

\section{Additional Results on the preparation of the 1D MPS Family}

\subsection{Scaling of the Fidelity with system size $N$}
It was shown in Ref.~\cite{PhysRevResearch.5.L022037} that, for quasi-adiabatic preparation of the MPS family with evolution time $T$ and system size $N$, the resulting fidelity exhibits the following scaling behavior [same as Eq.~(14) in the main text]:
\begin{equation}
\label{fid_supp}
\mathcal{F}(N,T) = \exp\left[-\kappa(T)\, N - c(T)\right],
\end{equation}
where $\kappa(T)$ denotes the error density and $c(T)$ is a boundary term that does not scale with the system size $N$. 

This scaling reflects the nature of quasi-adiabatic evolution: residual excitations generated by the finite evolution time act as a finite density of ``errors,'' leading to an exponential decay of the fidelity with $N$. Since counterdiabatic (CD) driving suppresses the density of residual excitations throughout the system, we expect the same scaling form to hold for the counterdiabatic evolution at finite time $T$.

In Fig.~\ref{fig:exponential_decay}, we show the scaling of the fidelity $\mathcal{F}$ with system size $N$ for preparing a 1D MPS with correlation length $\xi \approx 2.5$, at various fixed evolution times $T$, using (a) adiabatic evolution and (b) counterdiabatic (CD) driving with the first-order WNC ansatz. For each fixed $T$, we find that the fidelity $\mathcal{F}(N,T)$ decays exponentially with $N$. As $T$ increases, the slope of the exponential decay decreases, indicating that longer evolution times reduce the error density.

This behavior confirms that the scaling form in \cref{fid_supp} also applies to the counterdiabatic dynamics. From these data, we extract the error density $\kappa(T)$ and the boundary term $c(T)$, as shown in Fig.~3(a,b) of the main text. Furthermore, Fig.~3(c,d) in the main text demonstrates that the scaling form \cref{fid_supp}, with $\kappa(T)$ and $c(T)$ obtained from the finite-size scaling data (cf.~Fig.~\ref{fig:exponential_decay}) up to $N=100$, accurately predicts the required evolution time $T_p$ to reach a fixed target fidelity. The prediction agrees with TDVP simulation results up to $N=1000$, supporting the validity of the scaling form $\mathcal{F}(N,T)$ for large many-body systems.

\subsection{More Results for different correlation lengths, and the boundary term $c(T)$}

\begin{figure*}[h!]
\centering
\includegraphics[width=12.5cm]{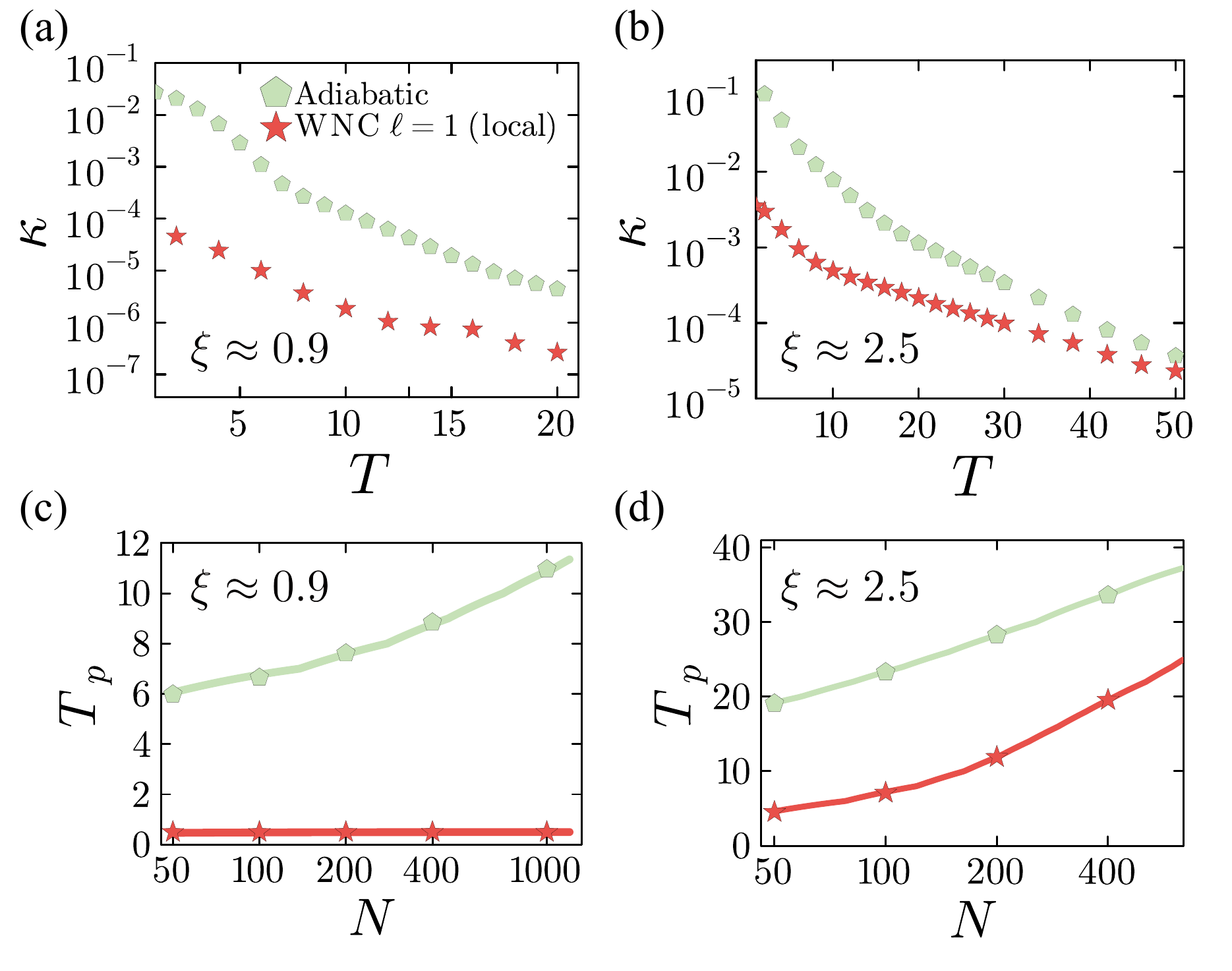}
\caption{Preparation of the MPS family with correlation lengths $\xi \approx 0.9$ and $2.5$. 
(a,b) Extracted error density $\kappa(T)$ [cf.~\cref{fid_supp}] as a function of the evolution time $T$. 
(c,d) Required runtime $T_p$ to reach the target fidelity $\mathcal{F} = 0.95$ as a function of system size $N$. 
Solid green and red lines indicate the predictions from \cref{fid_supp}, while the symbols denote TDVP simulation results.
 }
\label{fig:kappa}
\end{figure*}

In Fig.~\ref{fig:kappa}, we present the scaling of the error density $\kappa(T)$ and the required evolution time $T_p$ to reach a target fidelity $\mathcal{F} = 0.95$ for preparing the MPS family with correlation lengths $\xi \approx 0.9$ and $\xi \approx 2.5$. These results complement Fig.~3 in the main text, which shows the cases $\xi \approx 1.6$ and $\xi \approx 3.8$. 

In panels (a,b), we observe that CD driving based on the first-order WNC ansatz significantly reduces the error density $\kappa(T)$. For the small correlation length case $\xi \approx 0.9$, $\kappa(T)$ at short evolution times is reduced by nearly three orders of magnitude. This substantial suppression of the error density translates into a markedly reduced preparation time $T_p$ for system sizes up to $N=1000$. 

Panels (c,d) further demonstrate that the WNC approach significantly lowers the required evolution time $T_p$. Notably, for $\xi \approx 0.9$, the WNC protocol yields such a small error density that the required time $T_p$ shows almost no scaling with $N$ up to $N=1000$ in order to reach the target fidelity $\mathcal{F} = 0.95$. Moreover, the TDVP simulation results are in good agreement with the $T_p$ predicted from the scaling form \cref{fid_supp}.

\begin{figure*}[h!]
\centering
\includegraphics[width=12.5cm]{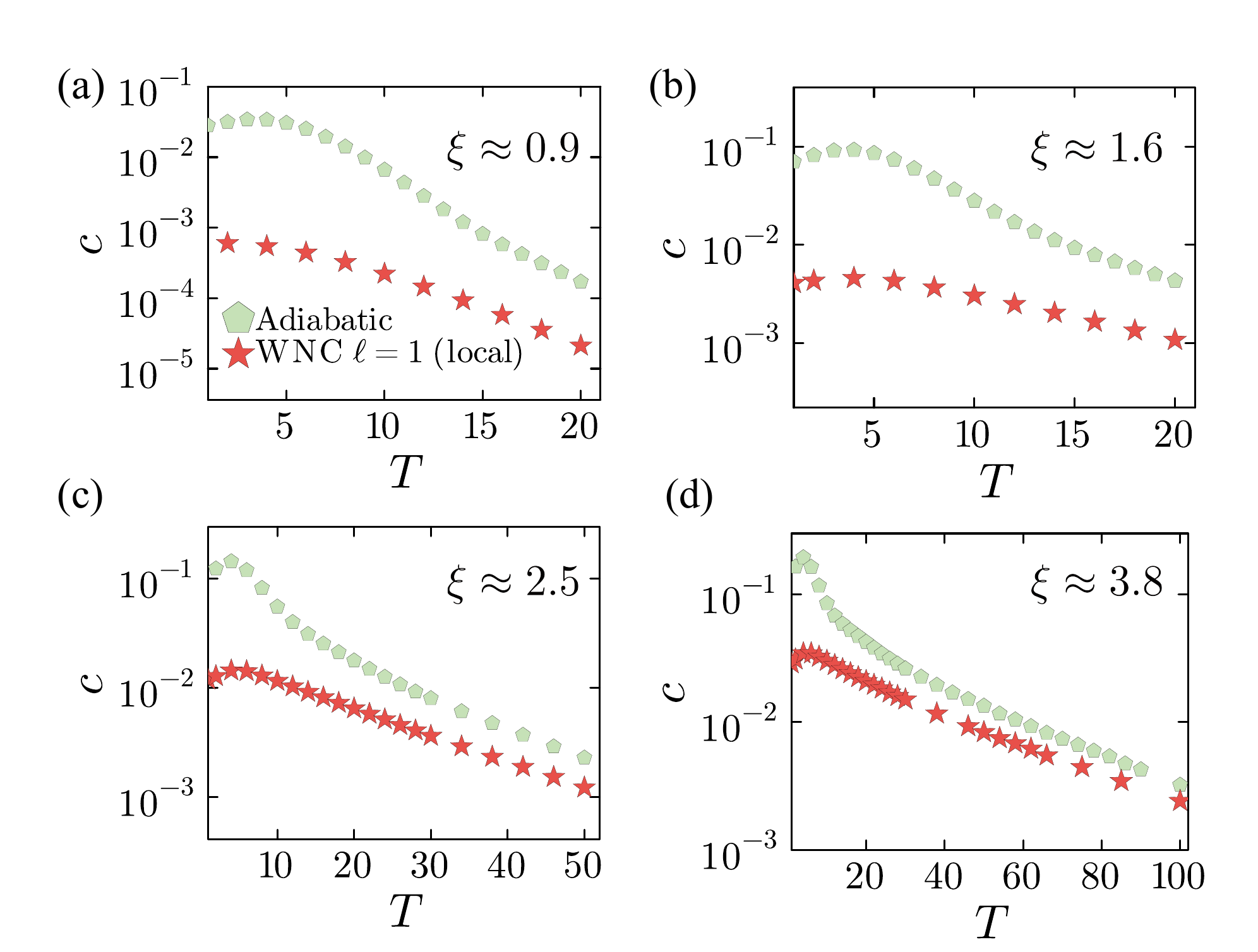}
\caption{Extracted boundary term $c(T)$ [cf.~\cref{fid_supp}]  versus the evolution time $T$ for the preparation of the MPS family with correlation lengths $\xi\approx 0.9,1.6,2.5,3.8$. 
 }
\label{fig:c_g}
\end{figure*}

Finally, in Fig.~\ref{fig:c_g}, we plot the boundary term $c(T)$ for both adiabatic evolution and the first-order WNC protocol at different correlation lengths $\xi \approx 0.9, 1.6, 2.5,$ and $3.8$. We observe that $c(T)$ remains small over the entire range of evolution times $T$. Moreover, the WNC approach consistently reduces this boundary error term compared to adiabatic evolution at each fixed $T$.

\subsection{Comparing gate costs in Trotterized simulations of various state-preparation schemes}
As shown in Fig.~2 of the main text, the first-order WNC ansatz already achieves a lower state-preparation infidelity than high-order NC ansatz. Moreover, implementing the first-order WNC ansatz experimentally is expected to be substantially easier than realizing high-order NC ansatz. To make this point explicit, we evaluate the gate cost of the digital Hamiltonian simulation of the Hamiltonian paths $H[s(t)]$ (adiabatic) and
$H_{\rm CD}[s(t)]$ (CD, for various ansatze, including first-order WNC and high-order NC) [cf.~Eqs.~(1) and (2) in the main
text]. 

For the time-dependent Hamiltonian, we discretize the evolution into $N_{\rm steps}$ intervals of duration $\tau$, such that the total evolution time is $T=N_{\rm steps}\tau$. The first-order product-formula approximation is defined as
\begin{equation}
U_{\rm Trotter}(T)=
\prod_{k=1}^{N_{\rm steps}}
\prod_{j=1}^{N_h}
e^{-i\tau h_{j}[s(k\tau)]},
\label{trotter}
\end{equation}
where $h_{j}[s(k\tau)]$ denotes the $j$-th local term of the Hamiltonian at time $k\tau$. Applying this Trotterized evolution to the initial state $|\psi(0)\rangle$, we obtain the Trotterized final state $|\phi’(T)\rangle$ [cf.~near Eq.~(13) in the main text]. We then use the fidelity $\mathcal{F}_{\rm Trotter}=|\langle\psi(1)|\phi’(T)\rangle|^2$ to quantify the performance of the digital state preparation. Note that Trotter errors are expected to reduce $\mathcal{F}_{\rm Trotter}$ and are therefore already taken into account in our analysis.

We focus on estimating the gate cost of preparing the MPS family. Because the local terms of the MPS parent Hamiltonian [Eq.(13) in the main text] typically do not admit a simple Pauli-string decomposition, we treat each $e^{-i\tau h_j[s(k\tau)]}$ as a generic $m$-qubit unitary, where $m$ denotes the support size of the local term $h_j(s)$. Implementing such a generic $m$-qubit unitary requires at most $N_{\rm CNOT}^{(m)} = \lceil \frac{1}{4}(4^m - 3m - 1)\rceil$ CNOT gates~\cite{PhysRevA.93.032318}, which we use as our estimate of the CNOT cost for each unitary. The total CNOT gate cost is then obtained by summing the costs of all unitaries within $U_{\rm Trotter}(T)$ [cf.~Eq.~\eqref{trotter}].

\begin{figure*}[t]
\centering
\includegraphics[width=10cm]{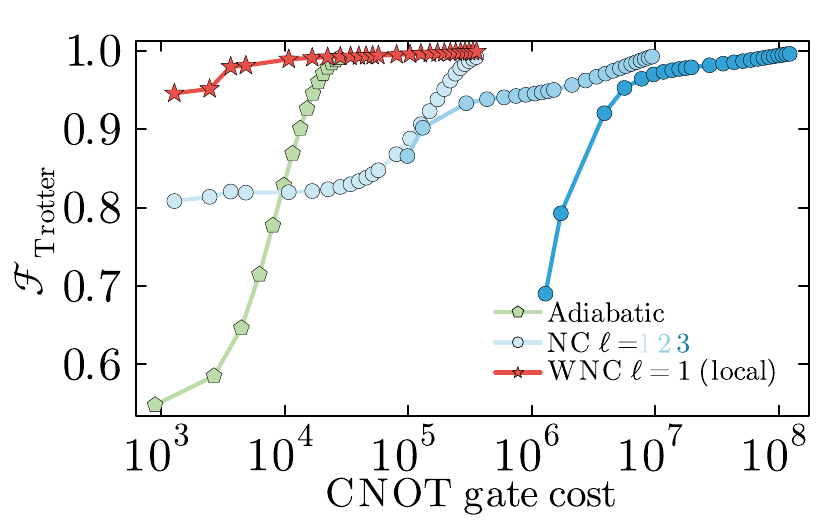}
\caption{Preparation of the 1D MPS family with correlation length $\xi \approx 3.8$, system size $N=6$, using the adiabatic protocol and CD driving with various ansatz. 
Fidelity as a function of the total CNOT gate count using the first-order trotterized evolution [cf.~Eq.~\eqref{trotter}].}
\label{fig:appendxi2}
\end{figure*}

In Fig.~\ref{fig:appendxi2}, we show the fidelity $\mathcal{F}_{\rm Trotter}$ as a function of the CNOT gate cost for preparing the MPS family with correlation length $\xi \approx 3.8$ and system size $N=6$, using the schedule function $s(t)=\sin ^2\left[\frac{\pi}{2} \sin ^2\left(\frac{\pi t}{2 T}\right)\right]$. We fix the time step to $\tau=0.05$ and vary $N_{\rm step}$ to change the total evolution time $T$; accordingly, the CNOT gate cost also varies. The choice of a constant time step is motivated by current near-term quantum devices, where the Trotter step size is typically taken to be constant.

From Fig.~\ref{fig:appendxi2}, we see that at a relatively low CNOT gate cost of $\sim 10^3$, counterdiabatic driving based on the first-order WNC ansatz already achieves a high fidelity of $\mathcal{F}_{\rm Trotter}\approx 0.94$. By contrast, the adiabatic protocol requires roughly an order of magnitude more CNOT gates to reach a comparable fidelity. Moreover, the high-order NC ansatze require substantially more CNOT gates to achieve similar performance, and in fact perform even worse than the adiabatic protocol because of the large overhead associated with high-order nested commutators. These results therefore highlight that the first-order WNC ansatz not only substantially accelerates the adiabatic evolution, but also introduces only a modest overhead comparable to that of the first-order NC ansatz. As a result, it provides the most resource-efficient approach for preparing the target state to a given high fidelity.

\bibliography{bibfile}